\documentclass[%
 aip,
 amsmath,amssymb,
 reprint,%
]{revtex4-1}

\usepackage{graphicx}% Include figure files
\usepackage{dcolumn}% Align table columns on decimal point
\usepackage{bm}% bold math
\usepackage[utf8]{inputenc}
\usepackage[T1]{fontenc}
\usepackage{mathptmx}
\usepackage{etoolbox}

\makeatletter
\def\@email#1#2{%
 \endgroup
 \patchcmd{\titleblock@produce}
  {\frontmatter@RRAPformat}
  {\frontmatter@RRAPformat{\produce@RRAP{*#1\href{mailto:#2}{#2}}}\frontmatter@RRAPformat}
  {}{}
}%
\makeatother
\begin{document}

\preprint{AIP/123-QED}

\title{Evolutionary Multi-agent Reinforcement Learning in Group Social Dilemmas}
% Force line breaks with \\
\author{B. Mintz} 
\email{brian.a.mintz@dartmouth.edu.}
\author{F. Fu}%
\affiliation{ 
Mathematics Department, Dartmouth College%\\This line break forced with \textbackslash\textbackslash
}%

\date{\today}% It is always \today, today,

\begin{abstract}
    Reinforcement learning (RL) is a powerful machine learning technique that has been successfully applied to a wide variety of problems. However, it can be unpredictable and produce suboptimal results in complicated learning environments. This is especially true when multiple agents learn simultaneously, which creates a complex system that is often analytically intractable. Our work considers the fundamental framework of Q-learning in Public Goods Games, where RL individuals must work together to achieve a common goal. This setting allows us to study the tragedy of the commons and free rider effects in AI cooperation, an emerging field with potential to resolve challenging obstacles to the wider application of artificial intelligence. While this social dilemma has been mainly investigated through traditional and evolutionary game theory, our approach bridges the gap between these two by studying agents with an intermediate level of intelligence. Specifically, we consider the influence of learning parameters on cooperation levels in simulations and a limiting system of differential equations, as well as the effect of evolutionary pressures on exploration rate in both of these models. We find selection for higher and lower levels of exploration, as well as attracting values, and a condition that separates these in a restricted class of games. Our work enhances the theoretical understanding of evolutionary Q-learning, and extends our knowledge of the evolution of machine behavior in social dilemmas.  

    % Research question - does learning, or evolution, or their combination solve the social dilemma? Under what circumstances (reward functions). How can we understand evolution in these complex systems? e.g. plot payoff before and after evolution of Temperature for various reward functions. What did we learn - don't blindly trust the results of a genetic algorithm, the fixation probability approach can give you a more holistic understanding than a bunch of runs, but it limited to low muation rates and has difficulty with many params. By using analytic approaches that average out stochasticity, one may be able to optimize parameters more efficiently. 
\end{abstract}

\maketitle

\begin{quotation} % variation of abstract, "advertises the main points of the article. It must describe in terms accessible to the nonspecialist reader the context and significance of the research and the importance of the results"
This study applies a powerful, widely used artificial intelligence framework to a social dilemma to study the issue of AI coordination. By using an evolutionary approach, we expand the understanding of the complex dynamics that this system exhibits. 
\end{quotation}

\section{\label{sec:intro}Introduction:}

% Broad AI concerns
The world has recently seen a surge of innovations powered by advanced artificial intelligence technologies, such as Large Language Models and self-driving cars, promising to fundamentally reshape many aspects of the world. As progress in these continues, we will see such systems deployed more extensively throughout the world. Already this has created complex systems of interacting agents with different goals and patterns of behavior. Understanding the theoretical basis of these systems will be crucial for successful implementations, and require new approaches with interdisciplinary ideas \cite{tsvetkova2024new}. In particular, a pressing open question is how to ensure models act cooperatively while performing their given task well \cite{dafoe2021cooperative}. This is related to the issue of aligning the incentives used in training AI models with those of the broader society.

% The power of RL / MARL, applications and problems. 
Reinforcement learning (RL) is one prominent framework for AI that has been successfully applied to many challenging problems due to its exceptionally general approach \cite{kaelbling1996reinforcement}. Whereas many machine learning techniques have constraints on the types of problems they can address, this approach can be applied to a wide variety of problems. The key idea behind reinforcement learning is simply that actions that have a positive payoff will be repeated more, and those with a negative payoff will be repeated less \cite{watkins1992q, szepesvari2022algorithms}. This model has its basis in models of animal psychology, but has since found a series of cutting-edge applications due to its flexibility. Reinforcement learning has been applied to a wide variety of problems from traditional games like Go, to stock price predictions, the management of energy systems, and controlling chaotic dynamics \cite{barto2017some, perera2021applications, bucci2019control}. This approach can be unpredictable, especially when multiple agents learn simultaneously, as this creates a dynamic environment \cite{leonardos22exploration, kianercy2012dynamics, nguyen2020deep, bloembergen15evolutionary, bowling2002multiagent}. Such Multi-Agent Reinforcement Learning (MARL) systems have likewise seen many applications such as navigating groups of autonomous vehicles and distributing resources through communication systems \cite{gronauer2022multi, du2021survey}. This field draws attention from researchers studying complex systems, engineers seeking to improve algorithms, and policy makers trying to manage these technologies effectively. % Many complex dynamics can be seen even for simple models, such as chaos citation. For example, \cite{leonardos22exploration} discusses how one can understand the path dependent results through the geometry of the Quantal Response Equilibria surface. % This is multiagent? % There are also several modifications of this algorithm, discussed in the survey of \cite{bloembergen15evolutionary}. % this has been a main focus of previous work, modifying it for better performance in specialized cases, a lot of offshoots, little theoretical understanding, we're adding to this. %\cite{leonardos22exploration} shows that an analogue of Q-learning has bounded regret, converges to the Quantal Response Equilibria, a classical solution with bounded rationality. Links findings to catastrophe theory, phase transitions in system equilibria given the exploration parameter. Potential games.  For example, Q-learning is proven to converge to the optimal choice of action in each state given sufficiently many update steps, and a decreasing learning rate \cite{watkins1992q}.

% Evolutionary MARL. We choose PGG becasue it studies the cooperation issue is general enough to capture a broad range of interactions. 
Recently, there has been growing interest in combining reinforcement learning with evolutionary algorithms, a similarly general approach with simple motivation. Evolutionary algorithms use mutation and selection to solve complex problems \cite{miikkulainen2021biological, li2023machine, tang1996genetic, haldurai2016study}. Reinforcement learning has a natural connection to this method, as learning in a single agents is equivalent to evolutionary dynamics between the strategies that agent can use with a particular form of mutation \cite{tuyls2003selection, kianercy2012dynamics}. Thus evolution between agents themselves can be seen as a form of multi-level selection, which has long been investigated by evolutionary theorists. The majority of the work in this intersection focuses on comparing and improving different algorithms, as opposed to understanding its theory \cite{sehgal2019deep, dominic1991genetic, moriarty1999evolutionary, liu2009study, wu2024evolutionary, kim2023evolving, bai2023evolutionary}. Consequently, a wide variety of MARL architectures have been proposed, with studies determining when each is optimal for various applications \cite{zhu2023survey, lin2024evolutionary, tan2021differential, liu2020mapper}.

% our contributions / approach - combining learning and evolutionary dynamics
Our work seeks to expand the theoretical understanding of these evolutionary MARL systems, eventually allowing a more principled approach to choosing or designing algorithms for MARL. Specifically, we investigate the effect of selection on a parameter governing the degree of exploration agents follow, and how this depends on the game determining agent interactions. Due to its relative analytic tractability, we focus on one of the foundational reinforcement learning models, Q-learning.  We combine this with a variety of ideas from Evolutionary theory to study dynamics within this system. To investigate cooperative AI, we focus on the evolution of learning in a classical social dilemma, the public goods game \cite{szolnoki2012conditional, santos2008social}. These are a canonical example of conflicting incentives, the interests of the self and those of the collective, and are suitably broad to encompass a wide variety of scenarios. % highlight this is more "multi" than most, which use 2 for tractability. 

% Prior work
Most prior work on evolution in social dilemmas have studied simple, often static, strategies \cite{hauert2006evolutionary, kurokawa2009emergence, mintz2023social, mintz2022point}. This is extended to more realistic agents in our work. Likewise, evolutionary MARL generalizes techniques like particle swarm optimization and simulated annealing to explore more efficiently \cite{wang2018particle, song2004research}. Previous study into reinforcement learning in social dilemmas has mainly focused on the Prisoner's Dilemma \cite{barfuss2023intrinsic, sandholm1996multiagent, izquierdo2008reinforcement, kaisers2011faq, masuda2011numerical, macy02learning, wang2022levy, yu2015emotional, oroojlooy2023review}. However this game is limited to interactions between two individuals, as with most theoretical studies of MARL, limiting the complexity that can be found but also reducing the potential applicability of the findings. A few studies have considered MARL in the public goods game, investigating the effect of learning models and how to optimize cooperation \cite{leimar2019learning, nax2015directional, wang2024enhancing, wang2023synergistic}. The majority look to use MARL to explain test subjects' behavior in experiments, \cite{iwasaki2003does, horita2017reinforcement, janssen2006learning, arifovic2004scaling, buhren2023social, cotla2015learning, hichri2007emergence}. Our study extends these preceding works by revealing a rich variety of evolutionary dynamics between learning agents in a group social dilemma. 

\section{Model}
% summary / overview of the two models
In this work, we expand the MARL framework to include evolutionary dynamics between agents. Inspired by evolutionary game theory, we investigate the population dynamics in a model where agents reproduce and die concurrently with learning. We study this through extensive agent based simulations of a stochastic model, and an evolutionary analysis of a corresponding deterministic system of ordinary differential equations. % These complementary approaches provide different perspectives on the dynamics in this model. 

% studying the evolution of a key parameter how several parameters effect pro-social behavior

% Reinforcement learning model
Our model uses the foundational Q-learning algorithm to perform reinforcement learning. In general, each agent keeps a table of values for each possible state of the system and possible choice of action. These are updated according to \begin{equation}
    Q(s_{t+1}, a_{t+1}) = Q(s_t, a_t) + \alpha \left[r_{t-1} + \gamma \max_{a} Q(s_{t+1},a) - Q(s_t,a_t)\right]
\end{equation} where $s_t$ and $a_t$ are the state / action at time $t$, $\alpha$ is the learning rate 
that determines how quickly values are updated, $r_{t-1}$ is the reward received, and $\gamma$ is the discount rate governing the extent the agent cares about future rewards. In essence, this generalizes the notion of keeping a weighted average of the rewards for a particular action, and also considers the change in state an action will cause. Actions can then be selected in a variety of ways, we use the Boltzmann function to determine policies, as it is better suited to mathematical analysis. With this method, actions are selected at random proportionally to the exponential of their payoff divided by a parameter $T$, the ``temperature'' of the agent, which controls the exploration of new strategies by the agent. As $T$ approaches zero, only the action with the highest Q-value will be selected, corresponding to purely exploiting this best-known strategies. Temperature can vary with time or payoff relative to some aspiration level, though it can also be a constant value. We focus on stateless $Q$-learning, as the variable group composition makes it challenging to specify meaningful states.

% Thus, the strategy is $x_i = \frac{e^{q_i/T}}{\sum_j e^{q_j/T}}$ or in the case of simply two actions, $x = \frac{e^{q_0/T}}{e^{q_0/T}+e^{q_1/T}} = \frac{1}{1+e^{(q_1-q_0)/T}}$, a function of the difference.

% Watkins and Dayan, '92 prove this converges to the optimal policy given sufficient updates for each state/action pair and \alpha \to 0, also Kaisers & Tulys 2011; Kianercy & Galstyan 2012 show convergence of FAQ to Nash Equilibria for two player two action normal form games given a decreasing temperature, so the question really becomes about equilibrium selection, that is, there isn't hope for the PD, where the only equilbrium is pure defection. Similarly, HD probably wouldn't be interesting, since there is one equilibrium, and in SH, there is two, but it mainly depends on intial conditions. 

% The game, maybe move to intro to talk about previous results, same for above? Ask Feng. 
In this study, rewards are determined by the public goods game. This is a natural setting for arbitrary numbers of agents to interact, and has a long history of being used to understand group social dilemmas like the tragedy of the commons and free rider effects. Each agent, of a group of $N$, chooses whether to pay a cost of one to contribute, or not contribute, referred to as defection. The total contribution is then scaled by a reward function $f(x)$ and redistributed evenly. That is, the payoff of individual $j$ is \begin{equation}
    \pi_j(c_1,...c_N) = \frac{1}{N}f\left(\sum_{i=1}^N c_i\right)-c_i
\end{equation} where $c_i$ is the contribution, zero or one, of individual $i$. Classically, the reward function can be linear $f(x) = rx$ with $1 <r < N$, or contain more complicated nonlinear effects. For this work, we include the division by $N$ in the definition of $f(x)$ to more easily interpret it as the reward per individual, simplifying comparison between values of $N$. We'll represent these functions at the possible discrete levels of contribution with vectors $[f(0), f(1), f(2), f(3), ..., f(N)]$, since the intermediate values cannot be realized so are irrelevant. This allows for more precise control of the rewards, though it has the drawback of being harder to extend to larger groups. 

% strategies could condition on other information about the co-players. 

%For example, if $k=3$, this is \begin{equation}
%    \pi(x) = (a - 3b + 3c - d)x^2+ (-2a + 4b - 2c)x + (a - b + 1)
%\end{equation} where $a, b, c, d$ are the payoffs if zero, one, two, or all three contribute. Note in general it appears the pattern uses the rows of pascals triangle for the coefficients (this is probably easy to prove based on the definition). There is a more general version of this formula in the python notebook, any collection of strategies (notation is a bit more opaque). 

\begin{figure*} % the * makes it two wide. 
    \centering
    \includegraphics[width=0.9\linewidth]{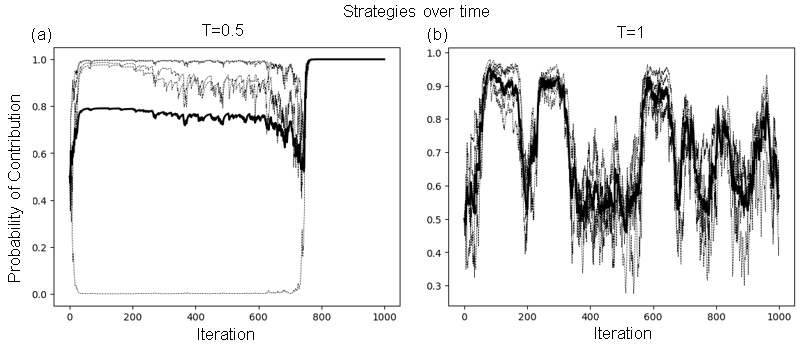} 
    \caption{\textbf{Stochastic learning dynamics with symmetric temperature}. These plots show the trajectories of strategies in the agent-based simulation over time as dotted lines, with the average strategy in bold, where the rewards are [0, 0, 0, 2, 4, 6], $N=5$, $\gamma = 0$, $\alpha = 0.1$, $r = 0$, and $T=0.5$ in panel (a) and $T=1$ in panel (b). By varying the temperature, a range of behaviors are possible. For low temperatures, relative to learning rate and rewards, agents enter a self-reinforcing cycle where they choose the most beneficial action repeatedly. For large temperatures, the strategies fail to converge. We see good alignment with the predictions of the ODE model, that strategies cluster together when the temperatures are the same. } % does it make sense for this reward function that learning, or are we just seeing drift?
    \label{fig:simulation}
\end{figure*}

% Our agent based model
Our first approach is an agent based simulation of a variation on a classical stochastic model of evolution, the Moran (death-birth) process\cite{nowak2006evolutionary}. A finite number of $N$ individuals interact to receive payoffs. At each time step, every individual has an independent probability $r$ of dying, then being replaced by the offspring of another member of the population, possibly with some mutation. The individual who gives birth for this replacement is selected proportionally to their fitness $f_i = e^{\beta \pi_i}$, the exponential of their average payoff $\pi_i$ over all previous interactions. Using the exponential of fitness is a standard technique to avoid complications from negative fitness. The parameter $\beta$ gives the strength of selection, in this study we mainly consider $\beta = 1$ for simplicity. Concurrently with the replacement, individuals perform Q-learning under Boltzmann Selection, with shared parameters $T$, $\alpha$, and $\gamma$, to receive rewards from the public goods game with reward function $f(x)$. This function could be linear such as $f(x) = kx$, which is most commonly considered, or contain non-linear effects such as $f(x) = b_0 x + b_1 x^2$. Sample trajectories of this simulation are shown in Fig. \ref{fig:simulation}. Death occurs uniformly at random at each iteration of the classical Moran process, so on average individuals only learn for $N$ iterations before being replaced. Our probabilistic modification gives agents more time to learn, an expected $1/r$ iterations, since the independent death probability means the iterations individuals learn for follows a geometric distribution with parameter $r$. Intuitively, the replacement rate parameter $r$ balances between learning, when it is low, and evolution, when it is high. We can then use this to estimate the fixation probability that a mutant will replace the resident type. These determine the evolutionary trajectories under rare mutations. In particular, they can determine whether there will be positive or negative selection, or other effects like attractors and repellors of the dynamics. 

% not really, high r is randomness. The classical case mostly corresponds to $r = 1/N$ (not independent, and it's assured). Also allows for larger turnover in the population % Pseudo-code? probably would look weird. % original model had a larger population, we decided not to go with this, as it added another independed source of group change, by selecting different groups to interact, in addition to the birth-death. 

Complementing this agent based simulation, our second model applies the evolutionary technique of adaptive dynamics using a limiting system of differential equations\cite{dieckmann2002adaptive,diekmann2004beginner,chen2023outlearning,brannstrom2013hitchhiker}. This is described more fully in the Appendix. 

\section{Results}
\begin{figure*}
    \centering
    \includegraphics[width=0.8\linewidth]{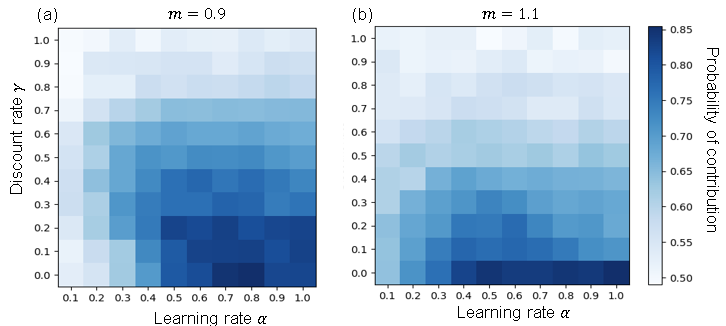}
    \caption{\textbf{The optimal learning parameters vary with the reward function}. This plot shows the average, over 100 runs, strategy in the group after 500 iterations where the horizontal axis is the learning rate and vertical axis is the discount factor, both between zero and one. Further, $r=0$ so there is no replacement, the temperature is $T=0.5$, the population consists of five agents, and the reward function is linear $f(x) =  kx$ with $k = 0.9$ on the left, and $1.1$ on the right. In these cases the jumps are a constant of $k$, so in the first it is always slightly better to defect, despite this agents contribute 85\% of the time with a higher learning rate and low discount rate. Similarly, on the right it is always slightly better to contribute, and a wider range of learning rates achieve a similar probability of contributing. Because of the nonzero temperature, it is impossible to achieve perfect cooperation, a strategy of one, and the achieved values are approximately the largest possible given this temperature and the possible rewards for each action. }
    \label{fig:learning-params}
\end{figure*}

\begin{figure}
    \centering
    \includegraphics[width=\linewidth]{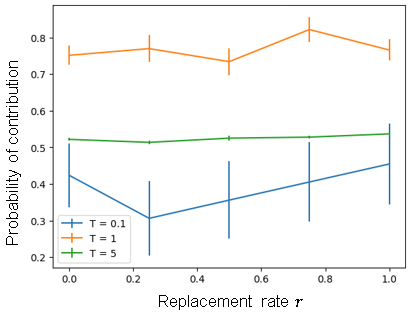}
    \caption{\textbf{Learning and evolution can have varying effects on contribution levels}. By varying the learning rate $T$ and replacement probability $r$, one can tune the relative strength of learning and evolution. This plots the average, plus or minus one standard error, over 20 runs of group level of contribution after 1000 iterations. These simulation are initialized with a single temperature, with no mutation in temperature, so selection is only acting on the strategies. } % can add a flipped plot, contrib vs T for various r, or various reward functions?
    \label{fig:strat-vs-param}
\end{figure}

%\subsection{Computational} % comment out before pre-printing. 
There is a significant effect from noise in the stochastic model, due to randomness in the action choices, which agents are selected to die, and which replace them. Because of this, careful consideration must be given to the model parameters to obtain meaningful results. In particular, if the rewards are large relative to the discount rate and temperature, it is possible that agents will prematurely fix their strategy and not explore sufficiently. Further, evolutionary dynamics are inherently more volatile in smaller populations, so larger groups or more trials are necessary. Controlling for these, we further account for the effect of the other parameters in this simulation by holding all others fixed before varying the parameter of interest, repeating this process for several combinations of the other parameters to ensure the results are robust. Figure \ref{fig:learning-params} plots the average probability of contribution depending on the learning rate and discount factor, showing a large learning rate and small discount factor are ideal in these cases. In other games, it is possible large discount factors or low learning rates are better. The two games shown use linear reward functions $f(x) = kx$ with $k=0.9$ and $1.1$ so the jumps in reward for an additional contribution are slightly above and below the cost of contribution of one, so learning should always favor contributing in the former case, and not contributing in the latter. Despite this, they can result in similar levels of cooperation, as the payoff is mostly determined by the actions of the other group members given the large group size. The combined effects of learning and evolution, given by the temperature and replacement rate parameters, are presented in Figure \ref{fig:strat-vs-param}. Depending on the temperature, we can see positive or neutral effects from replacement rate in a particular game. Across the replacement rates, we can see increasing temperature initially increases the contribution level, then decreases it to $0.5$ as agents purely explore. This suggests intermediate temperature values perform best, consistent with previous findings. Consequently, even within a single game, different effects from these forces are possible. 

% Larger groups tend to result in less cooperation, \ref{fig:boxplots}. This is consistent with the free rider effect observed in previous studies (citations). THIS IS BECAUSE THE JUMPS SHRINK IF YOU FIX THE FUNCTION. 

% Additionally, for small temperatures, we see agents split into pure cooperators and pure defectors. We believe this is because of the self reinforcing properties mentioned above. Even when the reward function strongly favors cooperation, for example if all the jumps are above one, some individuals can learn to defect. However, using the observation that learning favors contribution when its benefit exceeds its cost of one, we can design reward functions to promote specific behaviors. For example, if all jumps $f(k+1)-f(k)$ are above one, agents mostly learn to contribute, and if all are below one then defection is learned. By alternating large and small jumps we can create attractors and repellors of the dynamics. % is this all trivial / already known?

\begin{figure}
    \centering
    \includegraphics[width=\linewidth]{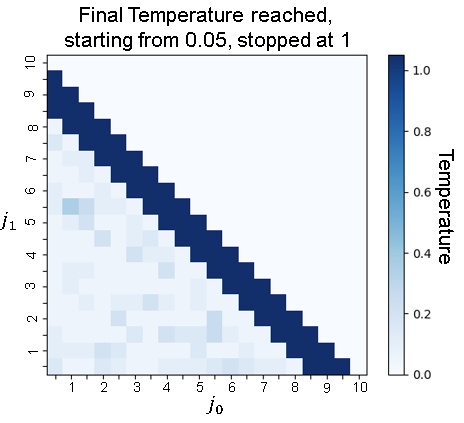}
    \caption{\textbf{The reward function can lead to positive or negative selection}. This plot represents the most likely outcome of the evolutionary dynamics in the temperature parameter, starting from $T=0.05$ and up to $T=1$, over the space of possible reward functions for the three player game, found through the adaptive dynamics approach described in the appendix. Letting $m$ be the maximum reward for when all individuals contribute, we can specify the function as $[0, j_0, j_0+j_1,m]$ where $j_0$ and $j_1$ are the jumps in reward when an additional individual contributes, if zero or one other had already contributed. Assuming the reward function is increasing, we have $0\le j_0\le m$ and $0\le j_1 \le m-j_0$, so only the values in the lower triangle are considered. We see there is a clear transition to larger final temperatures when $j_0+j_1$ exceeds a threshold depending on $m$, in this case $m=10$. } % the blurriness below could either be numerical instability due to the numeric solver, or intermediate levels selected for. I think it's the former. 
    \label{fig:reward-space}
\end{figure}

Our deterministic model reveals a wide range of possible selection effects on temperature for different reward functions. Specifically, we find examples where temperature experiences positive and negative selection, as well as attraction to an intermediate value. To understand how the game influences this, we study the direction of selection across all possible reward functions. Specifically, while the invasion fitness is positive, we repeatedly take mutant values slightly above resident values. This gives a sense for the most likely evolutionary trajectory of the temperature. Since there are $n+1$ values in the reward function for a group of $n$ players, we restrict to small cases to assist with finding patterns. In particular, we consider reward functions of the form $[0, j_0, j_0+j_1, m]$ where $0\le j_0 \le m$ and $0\le j_1\le m-j_0$. These correspond to a constraint where there is an upper limit $m$ on the per individual reward, and that the reward function is weakly increasing. Since there are only two parameters for a fixed $m$, we can plot the final temperature over the $j_0,j_1$-plane, shown in Figure \ref{fig:reward-space}. When $m > 3$ we see there is a clear separation between regimes, selection is positive when $j_0+j_1 = f(2)$ is above a threshold depending on $m$. For smaller values of $m$, selection is predominantly negative. The distinction is that for $m>3$, it is possible for all jumps to be above the cost of contribution one, where contribution would always be beneficial. This corresponds to the region $j_0>1$ and $j_1 > 1$, which does experience consistent selection on temperature. The computation of fixation probabilities generally supports these results, though computational constraints limited further investigation into these. 
% though this is not always the case.  The threshold cannot be theoretically determined, we might be able to plot it numerically

% Interestingly, there is some independence between the symmetric and full models: the payoff received when everyone has the same mutation parameter might suggest different things than when individuals with different parameters can interact. In particular, we can observe non-monotonic fitness with respect to a temperature parameter shared throughout the population, but a monotonic dynamics in that parameter when nearby invaders are considered. This means that the evolution does not simply maximize the group fitness. 

\section{Discussion}
% High level overview of the results / interpret / explain
Our results suggest that the relationship between learning parameters and cooperation levels is quite sophisticated, and warrants further study. The temperature of agents can evolve in a variety of ways depending on the environment. This suggests that combination of evolutionary algorithms with reinforcement learning can be used to optimize cooperation. Our approach of computing fixations probabilities presents and more interpretable than traditional evolutionary algorithms, providing a broader picture of the possible evolutionary trajectories that might be taken. Further, analytic approaches that average over possible interaction can improve the efficiency of these investigations and allow for more robust results through reducing stochastic effects. % adaptive systems could lose exploration or become totally random (the latter is surprising). What's the specific implicatiosn of THIS learning environment / game? 

% research question / how advances prior knowledge (not repeating below?) or specifically relate to other papers, e.g. our previous on MR. 

% What understanding did we add?

% Justify methods / limitations. 
While a variety of reinforcement models have been proposed, we focus on a the foundational Q-learning due to its theoretical guarantees and relative analytic tractability. Our analytic approach to studying these dynamics makes a few strong assumptions. We replace the reward an action receives with the average over all possible interactions, losing a large amount of data. Consequently, agents with the same temperature are expected to follow the same learning dynamics, often contrasting simulations where they can easily diverge based on initial actions. Future work could allow a more complete description of the possible states of the system, extending previous mean-field approaches that characterize the evolution of the probability distribution of strategies in the population \cite{galstyan2013continuous, yang2018mean, carmona2023model}. This would undoubtedly require a far more sophisticated mathematical model, and complicate the analysis. Another significant factor is the lack of states in our reinforcement learning model. This could simply be the number who cooperated in the previous round, or an average over the last several rounds. However this is a rather inaccurate measure of the true state, the strategies of all agents. % This would likely be a nasty PDE, as the dimension is the number of agents, and one has a joint distribution along this... Let alone if there are different 

% do we talk about the social implications, e.g. how evolutionary pressures on groups might have led to these learning properties?

\section{Conclusion}
% Background / Methods
In this work, we extended the MARL framework to allow for reproduction, introducing evolutionary pressures between agents. While previous studies have studied evolution on individuals with static strategies, or fixed groups with variable strategies, our model combines these two dynamics to create a more realistic system than either alone. By doing so, we are able to broaden the applications of evolutionary game theory, and bring theoretical insights into a complex learning model. We apply evolutionary techniques to study the dynamics in the temperature of agents, a key parameter governing their degree of exploration. In particular, we use agent based simulations to estimate the fixation probability of mutations in this parameters. These probabilities determine the evolutionary trajectories under the assumption of rare mutations. This assumption also allows us to apply adaptive dynamics using a system of ordinary differential equations to remove stochastic effects. 

% Results
Through extensive simulations, we explored the intricate relation between learning parameters and cooperation. In particular, temperature and replacement rate could have a variety of effects depending on other parameters like the reward function determining the type of public goods game being played. Depending on this, the temperature could evolve up or down, or to an intermediate levels. By studying a restricted class of these games, we conjectured a condition that determines which type of selection the temperature will undergo.  

% next steps
There are multiple possible extensions of this work. Firstly, this framework could be applied to a number of other games, such as the Iterated Prisoners Dilemma or coordination games. In particular, a round number of continuation probability could be used to determine how long the group interacts for, allow for more or less learning  to occur. By restricting to simpler cases, like small population sizes, we could derive further theoretical results. For example, the Moran process we simulated in model one could be explicitly represented to analytically determine the fixation probabilities. While the analytical results are often limited to simple cases, we could extend the simulations to capture more complicated effects, for example by including mutation in the parameter values. We could also consider a model where the initial strategy is also genetically determined, which would allow a more effective comparison between learning agents with nonzero temperature, and those who never change their strategy. We could also investigate the full replicator dynamics using the system of differential equations to determine finesses. This would provide a clearer picture of the dynamics between individuals with different temperatures. Future study in this direction has the potential to greatly enhance our understanding of the complex dynamics in Multi-agent Reinforcement Learning.  

% Gap / applications
Our approach expands the theoretical understanding of combining genetic algorithms with reinforcement learning. While evolution can and has produced remarkable solutions to many challenging problems, some situations can be ill-suited to this approach. Our investigation highlights the fact that careful consideration must be given to minimize the stochastic effects that can overpower selection. Such understanding is crucial as we see a myriad of applications of reinforcement learning. Further, these results provide initial implications for the coordination of AI systems, which are trained to optimize their own performance, yet must work successfully with other individuals. As such technologies continue to develop and connect with more aspects of our world, understanding the interfaces between models of different capacities will become increasingly important.

% This study avoided this because such results are less interpretable. 

% Payoff accumulation could be made geometric, not strictly the average (which weights old payoffs equally). 
% in the ODE model, integrate the payoffs rather than using the final value. This will penalize learning too slowly, even if the same end result is reached. Initial results indicate . 
% Use more sophisticated learning methods, such as Frequency-Adjusted Q-learning or WoLF. Could also extend to other Machine Learning techniques such as Neural Networks.
% could solve the ODE's with three or more exploration rates, this might not be too meaningful, but could show some new behaviors. 

\begin{acknowledgments}
B.M. is supported by a Dartmouth Fellowship.
\end{acknowledgments}

\section*{Data Availability Statement}
The data that support the findings of this study are available from the corresponding author upon reasonable request. % change to on my github later XXX, e.g. "The data that support the findings of this study are openly available in [repository name],  reference number [reference number].

\appendix*
% Deterministic Model and Adaptive Dynamics
\section{}
By assuming mutations are sufficiently rare, Adaptive Dynamics simplifies evolution to competition between two types: the resident, and a mutant which either fixates to become the new resident or becomes extinct before the next mutant emerges. Invasion fitness is the difference in payoffs $E(m,r)-E(r,r)$ between the mutant trait $m$ and resident trait $r$ when the mutant is initially rare, and is often used as a proxy for the fixation probability. Indeed, if this is negative, the mutant will experience negative selection and likely die out. Typically adaptive dynamics also assumes small levels of mutation to use the gradient to determine dynamics, but the lack of a closed form for our dynamics makes this infeasible, freeing up consideration to non-local mutation, as in the above model. Here, we assume interaction continues for long enough for the dynamics to reach equilibrium, and remains there long enough that the equilibrium payoff approximates the average payoff accumulated throughout the whole interaction, the quantity determining fitness in the first model. Since mutation is rare, we assume the group of $N$ consists of one individual having temperature $m$ while the others have temperature $r$, and find the equilibrium numerically by solving the system over a sufficiently long time range, estimated on a case by case basis. Specifically, we use the initial condition $x(0)=1/2$ since we assume the group forms without any prior information, so each agent initially follows a uniformly random strategy. Essentially, this approach separates the timescales between learning and evolution, performing selection based on the equilibrium reached. % because mutation is rare. 

\begin{figure}
    \centering
    \includegraphics[width=\linewidth]{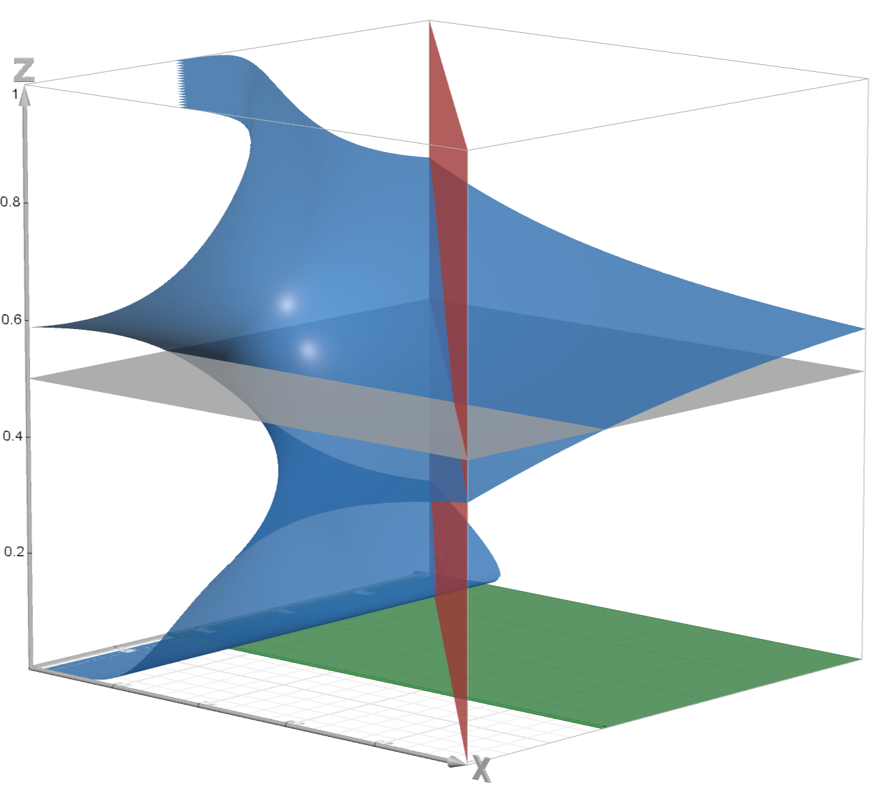} 
    \caption{\textbf{Null-manifold of symmetric learning dynamics over the space of three player Public Goods Games}. This plots the equilibria of the learning dynamics where $j_0=x$, $j_1=y$, and $z$ is the strategy, assuming $j_2 = m-j_0-j_1$. Specifically, this plot uses $m=3$ and $T=0.1$. The red plane delineates the rejoin $j_0+j_1 \le m$, and the green region is the subset where the initial rate of change of the strategy is positive. In this case, the maximum equilibrium contribution level, that is reached from an initial strategy of 0.5, occurs when $y$ is on this boundary, and $x$ is small. Note a large range of $x$, from zero to 0.1, have approximately the same level of contribution. Additionally, these values are close to having a negative initial change in the strategy, likely making them unstable for the actual dynamics, and thus possibly resulting in less frequent contribution. }
    \label{fig:manifold} % explain why this is being shown / what it demonstrates in the body. 
\end{figure}

% but what are the ODEs?
To derive our system of differential equations, we follow \cite{kianercy2012dynamics} and consider stateless Q-learning with no discounting ($\gamma = 0$), simplifying the Q-value update equation to \begin{equation}
    Q_i(t+1) = Q_i(t)+\alpha[r_i(t)-Q_i(t)] 
\end{equation} where $r_i(t)$ is the average reward of choosing action $i$ at time $t$, and $\alpha$ and $Q_i(t)$ are as before. If actions are chosen with the Boltzmann mechanism with temperature $T$, then the probability $x_i$ of choosing action $i$ is $x_i(t) = \frac{\exp(Q_i(t))/T}{\sum_i \exp(Q_i(t))/T}$. Taking the time derivative and rearranging and scaling time by $\alpha/T$, we find \begin{equation} 
    \frac{\dot{x}_i}{x_i} = \left[r_i - \sum_k x_k r_k \right] - T \sum_k x_k \ln \frac{x_i}{x_k} 
\end{equation} This first term corresponds to increasing the probability of choosing an action that has an above average payoff, and the second corresponds to the energy in a statistical-mechanical system. In fact, this shows that some reinforcement learning methods can be viewed as an evolutionary process within an agent between actions. When there are just two actions, we can summarize an agent's strategy by a single number $x$, the probability of choosing the first action. This gives the now single equation \begin{equation}
    \frac{\dot{x}}{x} = [r_1 - (x r_1 + (1-x) r_2)] - T \left(x \ln \frac{x}{x}+(1-x)\ln \frac{x}{1-x}\right) %= (1-x)[r_1 - r_2] - T (1-x)\ln \frac{x}{1-x}
\end{equation} or equivalently \begin{equation}\label{eqn:two-strat}
    \dot{x} = x(1-x)\left(r_1-r_2-T\ln\frac{x}{1-x}\right)
\end{equation} In the game we study the actions are contribution and defection, and the difference in expected payoffs is \begin{equation}
    r_1-r_2 = \sum_{S \subseteq \{1, ..., N-1\}} [f(|S|+1)-1-f(|S|)]\prod_{i\in S}x_i \prod_{i \notin S} (1-x_i)
\end{equation} the average difference in payoffs for contribution and defection over all possible subsets $S$ of individuals contributing, weighted by the probability of that outcome. These dynamics alone can exhibit a large degree of complexity, Fig. \ref{fig:manifold} depicts the loci of equilibria over the space of possible reward function for just three players, and a given temperature. Finally, the full system we consider has separate equations for $x_m$ and $x_r$, the strategies of those with the mutant and resident temperatures respectively, and the $x_i$ in $r_1-r_2$ are either $x_m$ or $x_r$ as appropriate. 

%\begin{align}
%    \frac{d}{dt} x_m &= x_m(1-x_m)\left(\left[\sum_{k=1}^{N-1} [f(k+1)-1-f(k)]\binom{N-1}{k}x_r^k(1-x_r)^{N-1-k}\right]-T_m\ln\frac{x_m}{1-x_m}\right) \\
%    \frac{d}{dt} x_r &= x_r(1-x_r)\left(\left[\sum_{k=1}^{N-2} R_k(x_m) \binom{N-2}{k}x_r^k(1-x_r)^{N-2-k}+(1-x_m)^{N-2-k} \right]-T_r\ln\frac{x_r}{1-x_r}\right) \\
%    R_k(x_m) &= x_m[f(k+2)-1-f(k+1)]+(1-x_m)[f(k+1)-1-f(k)]
%\end{align} where we simplified based on the other individuals possible strategies, for example the mutant only interacts with those having resident strategies so all $x_i = x_r$. % probably omit, not too pretty. Nice to show for shock value, joke "as you might imagine, this doesn't have a nice closed form equilibrium. 

\nocite{*}
\bibliography{MARL-refs}% Produces the bibliography via BibTeX.

\end{document}